\documentstyle[epsf,12pt]{article}
\newcommand{\insertplot}[1]{
\begin{center}\leavevmode\epsfysize=13.0cm \epsfbox{#1}\end{center}}

\newcommand{\be}{\begin{equation}}
\newcommand{\ee}{\end{equation}}
\newcommand{\ba}{\begin{eqnarray}}
\newcommand{\ea}{\end{eqnarray}}
\newcommand{\NL}{\nonumber \\}
\newcommand{\vs}{\vspace{0.5cm} }

\newcommand{\NN}[3]{ \left( N_{#1} N_{#2} - 
\frac{N^{eq}_{#1} N^{eq}_{#2} }{N^{eq}_{#3}} N_{#3}  
e^{({\Delta\mu_{#3}-\Delta\mu_{#1}-\Delta\mu_{#2} })/T} \right) }

\begin{document}
\baselineskip 14.5pt
\parindent=1cm
\parskip 3mm


\title{  Hadronization with a confining equation of state }

\author{ T.S. Bir\'o, P. L\'evai and J. Zim\'anyi \\ \\
      KFKI Research Institute for Particle and Nuclear Physics, \\ 
	 P. O. Box 49, Budapest, 1525, Hungary }
\date{September 16, 1998}
\maketitle

\begin{abstract}
We present a fast hadronization model for 
the constituent quark plasma ({\bf CQP}) 
produced in relativistic heavy ion collisions at SPS.
The model is based on rate equations and on a confining
equation of state inspired by the string phenomenology.
We display the time evolution of the relevant physical
quantities during the hadronization process and the
final hadron multiplicities. The results indicate that
the hadronization of {\bf CQP} is fast.
\end{abstract}



\section{INTRODUCTION}
\label{sec1}

The research of ultra-relativistic heavy ion collisions is from its
beginning motivated by the search for quark matter: a bulk of
deconfined color charges.
The prime attempts to describe the hadronization assumed, 
that this quark matter consists of non-interacting massless quarks 
and gluons \cite{Bir1,Raf,Knoll}.
In place of these early image of free quarks and gluons (plasma)
- which was based on thermodynamical studies of the pure nonabelian
gauge theory on the lattice -
gradually the picture of a quark matter emerges, containing effective
propagators and interaction vertices \cite{high-TQCD,Kamp,Lev1}.
At the characteristic energy scale of CERN SPS heavy ion
experiments ($\sqrt{s}/2 \approx 10$ GeV/nucleon) the dressed gluons
are heavier than the quarks 
($M_g \approx 600-800$ MeV, $M_q \approx 150-300$ MeV) \cite{Lev1}, 
and both values are 
bigger than the temperature ($T \approx 160$ MeV).
Therefore the number of quarks overweight that of the gluons and
can be treated non-relativistically.
On the ground of these theoretical indications we expect
that in the CERN heavy ion experiments not an ideal quark gluon plasma,
but a constituent quark plasma ({\bf CQP }), which contains antiquarks too,
 is formed in some
intermediate state of the reaction (see the ALCOR model,
Ref.~\cite{Bir2,Zim3}).

For the description of heavy ion reaction products at SPS
 a purely
hadronic interpretation of bulk experimental results
was also suggested
 \cite{Stach}. This alternative interpretation,
however,
has problems in explaining the process of the creation 
of new hadrons, especially the  short time scale.
Namely, the hadronic processes have characteristic
times of several
tens of fm/c \cite{Knoll,Zim1}, 
while the typical heavy ion reaction time at SPS energy
is about $1-2$ fm/c \cite{Bjorken}. 
Processes on the quark level on the
other hand have equilibration times of the order of
$0.1 - 2$ fm/c \cite{Bir1,Raf,Svetitsky+McLerran}.

The {\bf CQP}  picture has some advantages
in comparison with the quark-gluon plasma scenario too:
 In the hadronization of a quark-gluon plasma
the principles of color confinement and entropy generation
 can be both satisfied only with the assumption of an extreme
 growth of the reaction volume \cite{Zim2}. 
Further, a slow first order transition
 through near-equilibrium states
 would need too long time (up to $50$ fm/c),
 due to the re-heating \cite{Csernai+Kapusta}.  

In this paper we assume
that a massive quark matter, the  {\bf CQP}, is formed
in heavy ion reactions at SPS energy, which is
 in thermal but not necessarily in chemical
equilibrium.
The evolution of this {\bf CQP} is followed through a set
of coupled time dependent differential equations for 
the temperature and chemical composition. 
The construction of the  equations requires relations between 
the different equilibrium particle numbers at an actual temperature
as input.
 These input values are determined by the equation
of state. 

We mention in this place, that the ALCOR model 
\cite{Bir2,Zim3}, the hadronization problem is treated with 
an algebraic approximation, in contrast to the method
of the present paper, where we follow the complete time
evolution of the system. The inclusion of confining equation of state
in the present model is very important difference to the ALCOR model,
and to our best knowledge this type of phenomenological equation
of state with confining character has not been discussed in the
literature until now.

The paper is structured as follows: in Section \ref{sec_eos}. we discuss 
the equation of state of the mixture of hadrons
and interacting quarks. 
In Section \ref{sec_dyn}. we describe in details
the dynamics of the hadronization.
In Section \ref{sec_num}. we discuss our numerical results.
The conclusion is drawn in Section \ref{sec_con}.
In Appendix A. we collected the relevant hydrodynamical and
thermodynamical and statistical expressions used in the paper.

\section{EQUATION OF STATE OF THE MIXTURE OF HADRONS AND INTERACTING QUARKS}
\label{sec_eos}

We assume that at the beginning of the hadronization
the matter consists of massive quarks and anti-quarks.
In the time evolution of the system quarks and anti-quarks
form diquarks, anti-diquarks, mesons, baryons and anti-baryons.
We assume that the mixture of all of these particles are in thermal
equilibrium which can be characterized by a temperature.
For the representation of the interaction of the colored particles
we introduce an extra term into the free energy, which is inspired
by the string picture.

We note that there is an important difference to the canonical
approach to color confinement transition: in an (ideal) mixture of
quarks and hadrons the occupied volume, $V$, is the same for both
components, $V_q=V_h=V$,
the pressure
contributions $p_q$ and $p_h$ are additive. On the other hand
in the application
of the Gibbs criteria of a phase co-existence the volumes $V_q$ and
$V_h$ are additive $V_q+V_h={\rm constant}$, and the partial pressures 
are equal, $p_q=p_h=p$, in phase-equilibrium. 
In our physical picture of hadronization there is
no phase coexistence and the Gibbs criteria do not apply. Colored particles
and color-neutral clusters, pre-hadrons are distributed in a common
reaction volume and chemical reactions convert eventually 
the quark matter into a pure hadronic matter.

During this hadronization process the reaction volume
expands and cools.
Due to the change of the multi-particle composition this expansion
is not adiabatic: some heat can be produced (or consumed) by
quarkochemical processes. 
The expansion law of an ideal mixture 
follows from eq.(\ref{COOL}) and eq.(\ref{PRESSURE}) 
in Appendix A:
\be
\sum_i m_i \dot{N}_i + \frac{3}{2} T \sum_i \dot{N}_i
+ \frac{3}{2} \dot{T} \sum_i N_i + T \frac{\dot{V}}{V} \sum_i N_i = 0.
\label{STAR}
\ee
Due to the foregoing hadronization the number of particles decreases,
\be
\sum_i \dot{N}_i < 0,
\label{REHEAT}
\ee
therefore this process re-heats the system. Cooling effects are due to
the expansion \hbox{($\dot{V}/V=\partial_{\mu}u^{\mu} > 0$)} 
and rest mass creation 
\hbox{($\sum_i m_i \dot{N}_i > 0$).} The latter is possible with satisfying
(\ref{REHEAT}) only if hadron masses are  larger than
the sum of their constituents.

At the present level, however, the physical process of color
confinement by the hadronization is not yet taken into account.
In particular due to the dilution during the expansion
it  happens that
not all quarks or diquarks find a partner to hadronize. 
In order to avoid this effect we supplement the model
by the following confinement principle: 
all particles carrying color charge 
(quarks, diquarks, anti-quarks and anti-diquarks) will be penalized
by a free energy contribution stemming from strings.
The number of strings is proportional to 
a weighted sum of the number of color charges,
\be
Q = \sum_i q_i N_i,
\ee
Here it is $q_i=0$ for hadrons, 
$q_i = 1/2$ for quarks and anti-quarks, and $q_i = 3/4$ 
for diquarks and anti-diquarks. 
The higher effective charge of diquarks reflects a possibly higher
number of in-medium partners, to which a string is stretched.

The average length $L$ of a string  depends on the
density of colored objects, as $L=n_c^{-1/3}$, where
\be
n_c = \sum_{i \in c} N_i/V,
\label{eq_nc}
\ee
where the summation $i \in c$ excludes color neutral particles (hadrons).
So the free energy of the ideal quark matter - hadron matter mixture,
$F = F_{{\rm id}} + \Delta F,$
is supplemented by the following contribution of strings:
\be
\Delta F = \sigma_s n_c^{-1/3} Q,
\label{DELTA}
\ee
with the effective string tension $\sigma_s \approx 1.0 \  GeV/fm$.
We note here that if all colored particles carried the same color charge,
the effective string tension would be equal for all, and the interaction
free energy would become
\be
\Delta F = \sigma_{s, {\rm eff}} n_c^{2/3} V.
\ee
In order to achieve  baryon production and diquark elimination
properly we shall, however, need to use the more complicated ansatz
eq.(\ref{DELTA}).

This additional free energy comprises the non-ideality of the equation of
state we use. Since this addition is proportional to the volume $V$ and
the rest depends on densities only, it satisfies thermodynamical
consistency requirements \cite{Toneev} due to its construction.

While there is no new contribution to the entropy,
\be
S = S_{{\rm id}},
\label{ENTR}
\ee
the pressure, the energy and the chemical potentials of colored
($q_i \ne 0$) particles receive important modifications:
\ba
p     &=& p_{{\rm id}} - \frac{1}{3} \sigma_s n_c^{-1/3} \frac{Q}{V}, \NL
E     &=& E_{{\rm id}} + \sigma_s n_c^{-1/3} Q, \NL
\mu_i &=& \mu_{i,{\rm id}} + \sigma_s n_c^{-1/3} 
(q_i - \frac{1}{3} \overline{q}),
\label{NONID}
\ea
with $\overline{q} = Q/(Vn_c)$. Hadronic chemical potentials have no
modifications at all.

This non-ideal completion of the equation of state influences both
the expansion and cooling and the changes of particle composition.
Since the entropy is not changed by the introduction
of the interaction term (see. eq.~(\ref{ENTR})) therefore
\be
TdS = TdS_{{\rm id}} \ .
\ee
From this equation, using eq.~(\ref{PRINC}) we obtain
\be
dE + pdV - \sum_i \mu_i dN_i =
\left( dE + pdV \right)_{{\rm id}}
 - \sum_i \mu_{i,id} dN_i \ .
\ee
Applying eq.~(\ref{COOL}),
the non-ideal cooling law becomes
\be
 \left( dE + pdV \right)_{{\rm id}} +
\sum_i \left( \mu_i - \mu_{i, {\rm id}} \right) dN_i = 0 \ .
\label{NONCOOL}
\ee
Accordingly eq.(\ref{STAR}) is supplemented by a new generic term
due to the non-ideal equation of state,
\ba
\frac{\dot{T}}{T} &=& - \frac{2}{3} \frac{\dot{V}}{V} 
- \frac{\sum_i \dot{N}_i}{\sum_i N_i} 
- \frac{2}{3}\frac{\sum_i (m_i/T)  \dot{N}_i}{\sum_i N_i} \NL
&-&\frac{2}{3}\frac{\sum_i (\mu_i/T - \mu_{i,{\rm id}}/T)\dot{N}_i}{\sum_i N_i}
\ea
The additional term has the form
\be
\sum_i \frac{\mu_i- \mu_{i,id}}{T} \dot{N}_i =
\frac{\sigma_s n_c^{-1/3}}{T}
\sum_{i \in c} (q_i - \frac{1}{3}\overline{q}) \dot{N}_i \ .
\ee
This term is negative if color charges are eliminated from the mixture.
Therefore color confinement, causing an extra suppression of
equilibrium numbers of quarks and alike particles, re-heats the
expanding fireball as well as the ``normal'' chemistry of the
ideal quark - hadron mixture.
The more one suppresses color charges with respect to an ideal
mixture the more re-heating occurs during hadronization. This, of course,
works against the hadronization process.
The only physical effect besides a fast expansion - which has, however,
kinematical limits stemming from scaling relativistic expansion -
that can cool the mixture sufficiently is rest-mass production.
The color charge eliminating hadronization therefore must be
accompanied by the production of heavy hadron resonances.
(Rest mass production due to quark pair or gluon creation would not
reduce color.)
Since in our equation of state we have an explicit interaction energy
between the  quarks, our effective quark masses should be less
than that given in Ref.\cite{Lev1}. We shall use the following values:
$ ( m_u=[m_{u0}^2 + m_{th}^2]^{1/2} $,
$  m_d=[m_{d0}^2 + m_{th}^2]^{1/2} $,
$  m_s=[m_{s0}^2 + m_{th}^2]^{1/2} $,
with thermal mass $ m_{th} = 0.15 $ GeV and
$m_{u0}=m_{d0}\approx 0$, $m_{s0}=0.15$ GeV. 
The clusters have a mass according to the average mass
extra to the summed valence quark masses of the two lowest
lying hadron multiplets: the pseudoscalar and vector meson nonets
and the baryon octet and decuplet, respectively.

\section{DYNAMICS OF HADRONIZATION}
\label{sec_dyn}

\subsection{Initial state}
\label{subs_ini}

The initial energy density --- distributed along the beam direction
between $-\tau_0 \sinh \eta_0$ and  $\tau_0 \sinh \eta_0$ ---  can be related
to the center of mass bombarding energy $\sqrt{s}$ in the experiment,
\be
\varepsilon_0 = \frac{\sqrt{s}}{\pi R_0^2\tau_0 2 {\rm sh} \eta_0}.
\ee
On the other hand the initial invariant volume dual to $d\tau$
at constant $\tau=\tau_0$ is given by
\be
V_0 = \pi R_0^2\tau_0 2\eta_0.
\ee
The initial internal energy (i.e. the energy without the collective
flow of a fluid cell) at $\tau=\tau_0$ is therefore
less than $\sqrt{s}$ for finite $\eta_0$:
\be
E_0 = \varepsilon_0 V_0 = \frac{\eta_0}{ {\rm sh}\eta_0} \sqrt{s}.
\ee
At the CERN SPS experiment $\eta_0 \approx 1.75$ (due to some stopping),
$R_0 \approx 7$ fm, $\tau_0 \approx 0.8$ fm and
we obtain $V_0 \approx 431$ fm$^3$ and $E_0 \approx 2.13$ TeV.
Compared to the total energy of about $\sqrt{s} = 3.4$ TeV
(carried by about $390$ participating nucleons in a central
Pb-Pb collision)
approximately two third of the energy is invested into
rest mass of newly produced particles and thermal motion and
and one third  into the flow.

Comparing this with an alternative expression for the thermal energy
of an ideal massive quark matter,
\be
E_0 = \sum_i N_i(0) (m_i + \frac{3}{2} T),
\ee
one can estimate the initial temperature at the beginning of hadronization.
Using our standard values for the incoming quark numbers \cite{SQM97}
$N_u(0) = 544,$ $N_d(0) = 626$, further, assuming  that 400 $u \overline u$,
400 $ d \overline d $ and,  with $f_s=0.21 $, 
168 $ s \overline s$ quark anti-quark
 pairs are created in one central collision, 
 we get from the above equation $T_0 = 0.18$ GeV.
We use these numbers for the newly produced quark pairs 
in order to arrive the experimentally measured hadron and 
strange particle numbers.

\subsection{Hadronization processes}
\label{subs_hadr}

Chemical equilibrium is not supposed initially, rather a definite
over-saturation of quarks  in the reaction volume.
The initially missing color-neutral hadron states - mesons and
baryons - are formed due to quark fusion processes in a non-relativistic
Coulomb potential. The rates for different flavor compositions
differ mainly due to the different reduced masses of quark  anti-quark
or quark diquark pairs. First of all this influences the Bohr radius
in the Coulomb potential \cite{Bir2}. 
Of course, the presence of a medium -
which establishes the necessary momentum balance after the fusion -
also influences the hadronization rate. The cross section for
such a $2 \rightarrow 1$ process in medium is
\be
\sigma =  \left(\frac{\rho}{a}\right)^3
\frac{16M^2\sqrt{\pi}\alpha^2}{\left( \vec{p}^2 + 1/a^2 \right)^2}
\ee
with $a = 1/(\alpha m)$ Bohr radius of the $1s$ state in the
Coulomb potential and $\rho $ is the Debye screening length
\cite{Bir2}.  Here $\vec{p}$ is the relative momentum of the
hadronizing precursors, $m$ is their reduced mass and $M$ is the
total mass.

The coupling constant we are using is a function of the
relative momentum according to the formula
\be
\frac{1}{\alpha(p)} =  \frac{1}{\alpha_0} +
\beta_{{\rm QCD}} \log \frac{p^2}{\Lambda_{{\rm QCD}}^2}
\label{ALFA}
\ee
as long as $\alpha(p) < \alpha_0$. At smaller momenta
it levels off at $\alpha_0$.
Here
\be
\beta_{{\rm QCD}} = \frac{1}{4\pi} \left( \frac{11}{3} - 2N_F \right)
\ee
is the beta-parameter of the one loop beta-function
occurring in the charge renormalization process in perturbative
QCD. The function (\ref{ALFA}) is a good phenomenological approximation
and has the correct infrared
and ultraviolet asymptotics. The value $\alpha_0 \approx 1.4$
is taken in order to  fit  the gluon condensate strength
in vacuum.

The medium effect is parameterized in all cross sections 
by $\rho \approx 0.2 - 0.3$ fm in a flavor independent way. 
If the hadronization process is fast
enough, the prehadron numbers (i.e. hadron numbers before resonance decay)
become proportional to the two-body reaction rates. Since quark fusion
does not change the number of valence quarks during the
hadronization the fast hadronization limit leads to 
the ALCOR model \cite{Bir2}.

The relative momenta are taken from a random Gaussian
distribution,
\be
d{\cal P}(\vec{p}) \propto e^{-p^2/2mT} d^3p,
\ee
at temperature $T$ and reduced mass $m$.
This method allows us both to simulate thermally averaged
hadronization rates,
\be
R = \langle \sigma \frac{|\vec{p}|}{m} \rangle,
\label{RATE}
\ee
or to follow the event by event variation of the quark - hadron
composition. For a fast estimate of thermally averaged rates
about $3-5$ points in the relative phase space suffice for each reaction
at each time instant.

Although the above corrections to the equation of state 
help to reduce the equilibrium
ratio of free quarks to those confined in hadrons, total color
confinement would only occur after a long time when the mixture
is cool and dilute.
This is obviously not the case in relativistic
heavy ion collisions. Therefore we also apply a {\em dynamic confinement}
mechanism in our model: the medium screening length $\rho$
occurring in the hadronization cross section will be related to
strings pulled by color charges trying to leave the reaction zone.
This way the screening length $\rho$ is increased as the color
density decreases: we keep, however, the product $\rho^3 N_c/V$
constant,
\be
\rho(t) = \rho(0) * \left( N_c(0)/N_c(t) \right)^{1/3}.
\ee
Here $N_c$ stands for the number of colored objects
(each in triplet or anti-triplet representation),
\be
N_c = N_Q+N_{\overline{Q}}+N_D+N_{\overline{D}}.
\ee
As we shall discuss in the next chapter both color confinement
mechanisms are necessary in order to achieve a pure hadronic
composition in a short time by trans-chemical processes,
while satisfying the requirements of the entropy growth
and energy conservation.

\subsection{Reaction network}
\label{subs_net}

What remains to specify the model is the system of rate equations describing
the transformation of quark matter into hadronic matter.
We consider $N_F=3$ light quark flavors $u$, $d$ and $s$.

There are $N_F(N_F+1)/2=6$ possible diquark flavors and the same
number of anti-diquark flavors. The number of quark anti-quark 
flavor combinations is $N_F^2=9$ while that of quark or
anti-quark triplet combinations is $N_F(N_F+1)(N_F+2)/6=10$.
In the hadronizing quark matter we deal with altogether
$2*3+2*6+9+2*10=47$ sorts of particles.

Let us generally denote quarks by $Q$, diquarks by $D$,
mesons by $M$ and baryons by $B$.
The possible fusion reactions are:
\ba
Q + Q  & \longrightarrow & D,  \NL
\overline{Q} + \overline{Q}  & \longrightarrow & \overline{D},  \NL
Q + \overline{Q}  & \longrightarrow & M,  \NL
Q + D  & \longrightarrow & B,  \NL
\overline{Q} + \overline{D}  & \longrightarrow & \overline{B}. 
\ea
Accordingly the number of reaction channels, 
$2*10+9+2*6=41$ is $2N_F=6$ less than the number of
particle sorts. Therefore in chemical equilibrium,
when the above reactions are balanced
by the respective decays,
the number of all particles are under-determined.

This is fortunately not a problem, because
the hadronization reactions under consideration do not create or
annihilate elementary quark flavors, they just
produce new combinations.
There are therefore exactly $2N_F=6$ quark and anti-quark
flavor numbers, which are conserved by these reactions.
They are determined by the initial state.

Our model is completed by the system of rate equations.
Considering a general reaction of type
$$ i + j \longrightarrow k$$
we account for the changes
\be
dN_i = dN_j = - Adt,  \qquad  dN_k = +Adt,
\label{COMP}
\ee
cumulatively in each reaction.  Here
\be
A = R_{ij\rightarrow k}  N_i N_j 
\left( 1 - e^{\frac{\mu_k}{T} - \frac{\mu_j}{T} - \frac{\mu_i}{T} } \right)
\label{BACK}
\ee
with a thermally averaged  rate 
$ R_{ij\rightarrow k}$ (cf. eq.(\ref{RATE})).
The changes stemming from different reactions accumulate to a
total change of each particle sort in a time-step $dt$.

Inspecting the expression (\ref{BACK}) it is transparent that an extra 
increase of a chemical potential suppresses the production or enhances
the decay of the corresponding particle. This behavior ensures that
\be
\dot{S} = - \sum_i \frac{\mu_i}{T} \dot{N}_i > 0,
\ee
for any complicated network of reactions. The approach to chemical
equilibrium always produces entropy.

\subsection{Chemical equilibrium}

Chemical equilibrium is defined by the requirement that all chemical
rates vanish (cf. eq.\ref{BACK}). It leads to relations like
\be
  \mu_i^{{\rm eq}} + \mu_j^{{\rm eq}} = \mu_k^{{\rm eq}}
\label{CHEMEQ}
\ee
for each reaction channel. The correspondence between equilibrium
chemical potentials and equilibrium number densities is, however,
in the general case not as simple as for a mixture of ideal gases.
From eq.(\ref{NONID}) we obtain an implicit equation for the
equilibrium densities of colored particles,
\be
 n_i^{{\rm eq}} = n_{i}^{\rm th} \, 
\exp \left[ {\frac{\mu_{i}^{{\rm eq}}-b_i(n^{{\rm eq}}_c)-m_i}{T} } \right] \ ,
\label{eq_implic}
\ee
where $n_c^{{\rm eq}}$ is the color charge density in equilibrium,
(see Eq.(\ref{eq_nc})), $n_{i}^{\rm th}=N_{i}^{\rm th}/V$
from Eq.(\ref{MaxBolt}) and
\be
 b_i(n_c) =  
 \sigma_s n_c^{-1/3} 
 \left( q_i - \frac{1}{3} \overline{q} \right) \ .
\ee
We call the attention to the fact that no solution of Eq.(\ref{eq_implic})
exists below a critical temperature.

\vs
Since the number of reactions is less than the number of particle sorts,
the equilibrium state is not fully determined by these conditions alone.
The missing information is contained in the value of conserved numbers.

Applying eq.(\ref{NONID}),
the non-equilibrium chemical potentials, and hence the essential factors,
$e^{-\mu_i/T}$, in the chemical rates (eq.(\ref{BACK})) can be expressed
as
\be
 e^{-\mu_i/T} = e^{-\mu_i^{\rm eq}/T} \cdot 
e^{\frac{b_i(N^{{\rm eq}})-b_i(N)}{T}} \cdot
\frac{N_i^{{\rm eq}}}{N_i}.
\ee
In the combinations appearing in the detailed balance factor of the
rate equations using eq.(\ref{CHEMEQ}) we obtain
\be
 1 - e^{\frac{\mu_k-\mu_i-\mu_j}{T}} = 1 - 
\frac{N_i^{{\rm eq}}}{N_i} \frac{N_j^{{\rm eq}}}{N_j} \frac{N_k}{N_k^{{\rm eq}}}
e^{\frac{\Delta\mu_k-\Delta\mu_i-\Delta\mu_j}{T} }, 
\ee
with
\be
\Delta\mu_i = b_i(N) - b_i(N^{{\rm eq}}).
\ee
The corrections $b_i(N)$ in the non-equilibrium chemical potentials
may in general depend on the number densities of several other
components on the mixture.

\vs
At this point we note that the extra $e^{-\Delta\mu_i/T}$ factors
occur for non-ideal equations of state where the correction to
the free energy density is a nonlinear function of the number
densities.

\vs
\subsection{Detailed rate equations}

With the purpose of better understanding in mind 
we enumerate certain types of the rate
equations given in the above compressed form in eqs.(\ref{COMP},\ref{BACK}). 
Let $\{u, d, s, \ldots \}$ be an ordered set of flavor indices.
We denote 
   quarks by flavor indices $i,j$ or $k$, 
   the anti-quarks by their overlined versions,
   diquarks consisting of a flavor pair satisfying $i\le j$ by $[ij],$
   a meson made from a flavor $i$ quark and a flavor $\overline{j}$ 
   anti-quark by $[i\overline{j}],$
   and finally a baryon made from quark flavors $i, j$ and $k$ 
   satisfying $i\le j\le k$ by $[ijk]$. 
In fusion reactions we create primarily
mesons, (anti)diquarks and (anti)baryons.


The rate equations for these reactions 
are straightforward to encode on a computer but
lead to somewhat clumsy equations writing down the actual
rate equations: 
\ba
V \frac{dN_i}{dt} & = & - \, \sum_{j\ge i} 
R_{i+j \rightarrow [ij]} \NN{i}{j}{[ij]} \NL \NL
& - & R_{i+i \rightarrow [ii]} \NN{i}{i}{[ii]} \NL \NL
& - & \sum_{\overline{j}} 
R_{i+\overline{j} \rightarrow [i\overline{j}] } 
\NN{i}{\overline{j}}{[i\overline{j}]} \NL \NL
& - & \sum_{[jk]}
R_{i+[jk] \rightarrow [ijk]} \NN{i}{[jk]}{[ijk]}
\ea
\noindent for quarks,
\ba
V \frac{dN_{[ij]}}{dt}  & = &
 R_{i+j \rightarrow [ij]} \NN{i}{j}{[ij]}_{i \le j} \NL \NL
& - & \sum_k R_{[ij]+k \rightarrow [ijk]} \NN{[ij]}{k}{[ijk]}
\ea
for diquarks,
\be
V \frac{dN_{[i\overline{j}]}}{dt}  = 
R_{i+\overline{j} \rightarrow [i\overline{j}] } 
\NN{i}{\overline{j}}{[i\overline{j}]}
\ee
for mesons and finally 
\ba
V \frac{dN_{[ijk]}}{dt}  & = &
R_{i+[jk] \rightarrow [ijk]} \NN{i}{[jk]}{[ijk]}_{i\le j \le k} \NL \NL & + &
R_{j+[ik] \rightarrow [ijk]} \NN{j}{[ik]}{[ijk]}_{i < j \le k} \NL \NL & + &
R_{k+[ij] \rightarrow [ijk]} \NN{k}{[ij]}{[ijk]}_{i \le j < k} 
\ea
for baryons.
We obtain similar equations for anti-quarks, anti-diquarks and
anti-baryons.

\vs
We would like to point out that the novel appearance of the
above rate equations is a consequence of the equation of state
we are using. This form goes beyond the usual expression
multilinear in $N_i/N_i^{{\rm eq}}$, and reflects the fact that
the hadronization reactions are strongly influenced by the
presence of colored particles not taking part in the particular
two-body process directly.

\vs
Especially the rate equation for baryon production is complicated.
The conditions indicated as lower indices at the
bracket expressions or in the summation signs rule out multiple
counting of baryons which contain equal flavor quarks.
For example the reaction $u+[uu]\rightarrow[uuu]$ makes only
one $[uuu]$ baryon (pre-$\Delta^{++}$) and not three.
On the other hand $[uds]$ can be constructed from three
different quark - diquark fusion process
\ba
u + [ds] &\longrightarrow & [uds], \NL \NL
d + [us] &\longrightarrow & [uds], \NL \NL
s + [ud] &\longrightarrow & [uds]. 
\ea
The only easy and transparent notation we found for these
reactions is the compact eq.(\ref{COMP}).
Since we calculate the evolution of the number of a given type of quark
group (e.g. $uud$), at the final time we distribute this number between 
the corresponding multiplets ---
the lowest lying pseudoscalar and vector nonets for the mesons and 
lowest lying octet and decuplet for baryons (antibaryons) ---
according to the spin degeneracy. (E.g.
the number of $uud$ quark group is distributed between the $ p^+ $
and $ \Delta^+$ in the ratio two to four while $uuu$ populates only the 
 $\Delta^{++}$ resonance. )

\subsection{Hadronic decays}

The set of rate equations describes the time evolution of the
number of all involved particles. In order to get the final
hadron numbers we integrate these equations until the
number of colored particles becomes negligible. 
This way we obtain a number of hadronic resonances
(in the present version the vector meson nonet and baryon
decuplet).
Finally hadronic decays are taken into account with
the dominant
branching ratios obtained from Particle Data Table \cite{PDT}.
We assume that secondary hadron-hadron
interactions have a negligible effect on the
finally observed hadronic composition.

The time evolution of the entropy and temperature
is obtained by simultaneous integration of eqs.(\ref{ENTROPY})
and (\ref{STAR}). The structure of the rate equations
containing both creation and decay terms ensures 
that the entropy never decreases during the 
transition \cite{Thank_Knoll}.
On the other hand the energy conservation is established by
the effective cooling law eq.(\ref{STAR}).
The numerical study of this complex system of equations
is presented in the next section.


\section{ Numerical results and discussion}
\label{sec_num}

In this section we present the results of the numerical solution 
of the set of the $41+2$ coupled differential equations. In the 
Figures the  calculated time evolution of the temperature, 
entropy, internal energy and the
 different particle and antiparticle numbers are presented for a
158 GeV / nucleon Pb + Pb central collision. For the numerical solution
 the initial condition and the values of the parameters has to be
 specified. For the parameters describing the initial state we used those
given in subsection \ref{subs_ini}, while 
for the parameters determining the dynamics of the 
hadronization we used the
following values: $\rho = 0.2 fm$, $\alpha = 1.4$. 
Both the quarks and diquarks
vanished only in the case, if we used the value $ q_d = 1.5 $ for the
 effective diquark color string tension. The time
 evolution of different quantities 
calculated with this initial condition and parameters
 are shown in the Figures. 

From Fig.~1a we can see, that at the beginning of the hadronization 
there is a rapid decrease in the temperature due to the rest mass formation
of the hadrons. Shortly after that, the reheating starts as an effect
of color confinement (see eqs.(28) and (32)).
The eq.(28) shows, that by removing
two  colored object to produce a colorless hadron,
 the associated string energy is also removed, 
and it has to appear in the thermal energy. Since this energy is inversely 
proportional to the one third root of the color density, this 
effect is stronger at smaller color density.   Finally, as the
 hadronization is completed,
the expansion bleeds to the cooling of the system. Fig.~1b shows, that the 
total entropy is monotonically increasing during the hadronization,
while the contribution of colored particles (CQP) is gradually
eliminated.
In Fig.~1c one can
observe an interesting pattern in the time evolution of the pressure. 
The partial pressure 
of the interacting CQP rapidly decreases as the number of quarks 
decrease. As the color density drops, this pressure becomes even negative.
The increasing hadron partial pressure, however, overcompensates this negative
value. 
At later times with the expansion of the ideal hadron gas the total 
pressure decreases. Due to this interplay, the total pressure 
falls steeply in the early period, it stays for a short while
almost constant in the re-heating period, and finally drops slowly
at the late expansion stage.
Assuming an initial condition leading
to slower hadronization, this balance is not so effective and 
in the time evolution of pressure a local minimum appears. 
The partial and total internal energy evolution,
displayed in Fig.~1d, shows, that the hadronization is completed
at $2$ fm/c after the beginning of the process. The decrease of the 
internal energy is compensated by the 
work of pressure while making the flow (cf. eq.(\ref{COOL})),
the total energy density drops due to the volume expansion 
like $\epsilon \propto 1/\tau$ at very late times only,
when $p=0$.

Fig.~2 shows the time evolution of different colored particles. 
The quark and antiquarks numbers are monotonically decreasing functions
of time. The diquarks are produced from the quarks, and then they 
contribute rapidly to the formation of baryons. That keeps their number
always on a low level. In Fig.~3 the evolution of hadron  
numbers is shown. Here we did not take into account the decay  
 of the resonances. Actually these time evolutions are
the evolution of the constituent quark clusters, i.e. 
prime hadronic resonances, from which the experimentally
observed particles emerge.
 It is clear from this Figures, that the mesons are formed
faster than the baryons. This is understandable, if we consider, that
the mesons are produced in a one step process, while the baryons are
formed in a two step process: first the formation of diquarks and 
in a diquark quark reaction the formation of the baryon.
In the ALCOR model the ratio of hadronic species are determined
by the ratio of steepnesses of these curves. Since these curves do not 
cross each other, one can understand, why the algebraic ALCOR 
approach to the solution of rate equations is a good approximation. 
The difference, that at the very beginning the meson
curves increase linearly with the time and the baryon curves start
with a quadratic form, was taken into account in the ALCOR
model by one single common factor, the so called baryon 
suppression factor.  

In our model we calculate the number of all produced hadrons.
 In Table 1 the hadron numbers obtained with the Transchemistry
model and those obtained with the ALCOR \cite{SQM97}
and the RQMD \cite{RQMD} models are shown 
together with the few published  experimental data.
The present results of the ALCOR model are slightly different
from the previously published ones \cite{SQM97}, because here we
considered the feeding of $\Lambda^0$ 
(and $\overline{\Lambda^0}$) particle multiplicities from 
$\Sigma^0$, $\Xi^-$, $\Xi^0$, $\Omega^-$ 
(and $\overline{\Sigma^0}$,
$\overline{\Xi}^+$, $\overline{\Xi}^0$,
$\overline{\Omega}^+$
respectively), as it was obtained in the experiments.
The fit parameters to this new case are:
$N_{q{\overline q}}= 391$, $N_{s{\overline s}}= 172$
and $\alpha_{eff}=0.97$ comparing to the earlier
$N_{q{\overline q}}= 398$, $N_{s{\overline s}}= 175$
and $\alpha_{eff}=1.03$.

 Table 2 shows a comparison 
for the multi-strange baryon ratios. While in
many cases there are intriguing agreements, in some other cases 
 there are  some discrepancies.
The reason for this may originate from two sources: i) the
experimental data referred here are the production ratios in
 the overlap window of the detector acceptances.
Thus, if the momentum distribution of the two particle sorts
 is not the same,
then these ratios are not equal to the total number
 ratios. ii) these calculated
values are more sensitive to the simplifying assumption, that 
 the hadronization happens into
the lowest lying baryon octet and decuplet and the two lowest lying
meson nonet. We have committed this simplification, because the
inclusion of other hadron multiplets
would have multiplied the workload in the computer simulation.  
In our model we don't have such  parameters 
with which the different particle ratios could be manipulated
independently. The degree of the overall harmony between our results 
and the recent
experimental data seems to be promising and thus it could initiate 
the search for further signatures of the {\bf CQP} state.


\section{Conclusion}
\label{sec_con}

In this paper we presented a new model for the hadronization
of a constituent quark plasma ({\bf CQP}) based on rate equations in a
 quark matter - hadron matter mixture.
 The color confinement was taken into account by
using consistently a plausible equation of state 
motivated by the string model. The role of the different 
physical processes entering into the hadronization was discussed.
Our results presented in the Figures clearly show a very fast
hadronization.
This is mainly a consequence of the large hadronization
rates, $R_{ij \rightarrow k}$ and of the fact that in the initial state 
the system is very far from equilibrium.
Observing the shape of the time evolution of different
hadron multiplicities, it became understandable, why  
the simpler algebraic approximation, applied in the
ALCOR model, works so well. The comparison with the existing
experimental data indicate, that it is possible, that
 in the PbPb collision at SPS a piece of matter is formed, inside which
the massive quarks and anti-quarks interact with a string like
mean field.

Finally we emphasize, that  this type of phenomenological
 investigations
are necessary,  as long as the hadronization of the quark matter 
as a non-equilibrium, non-static,
non-perturbative process, cannot be described  with 
the methods of the elementary field theory. 

\insertplot{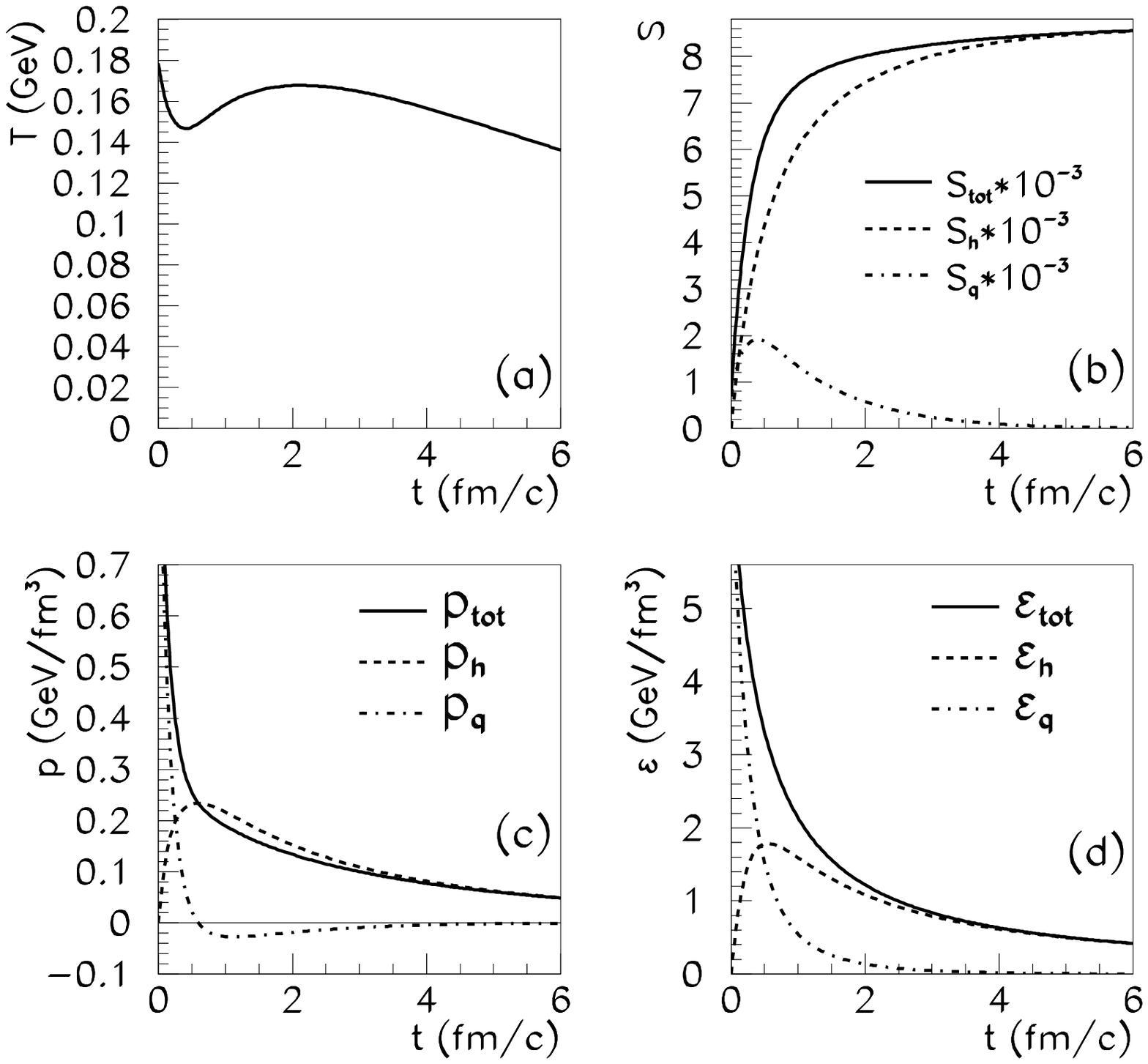}
 \begin{center}
  \begin{minipage}[t]{13.054cm}
  { {\bf Figure 1.}
The time evolution of the temperature $T$ (a), the
entropy $S$  (b),
the pressure $p$ (c) and the energy density $\varepsilon$ (d)
of the system together with the partial contributions of quarks and hadrons.}
  \end{minipage}
 \end{center}

\newpage

\insertplot{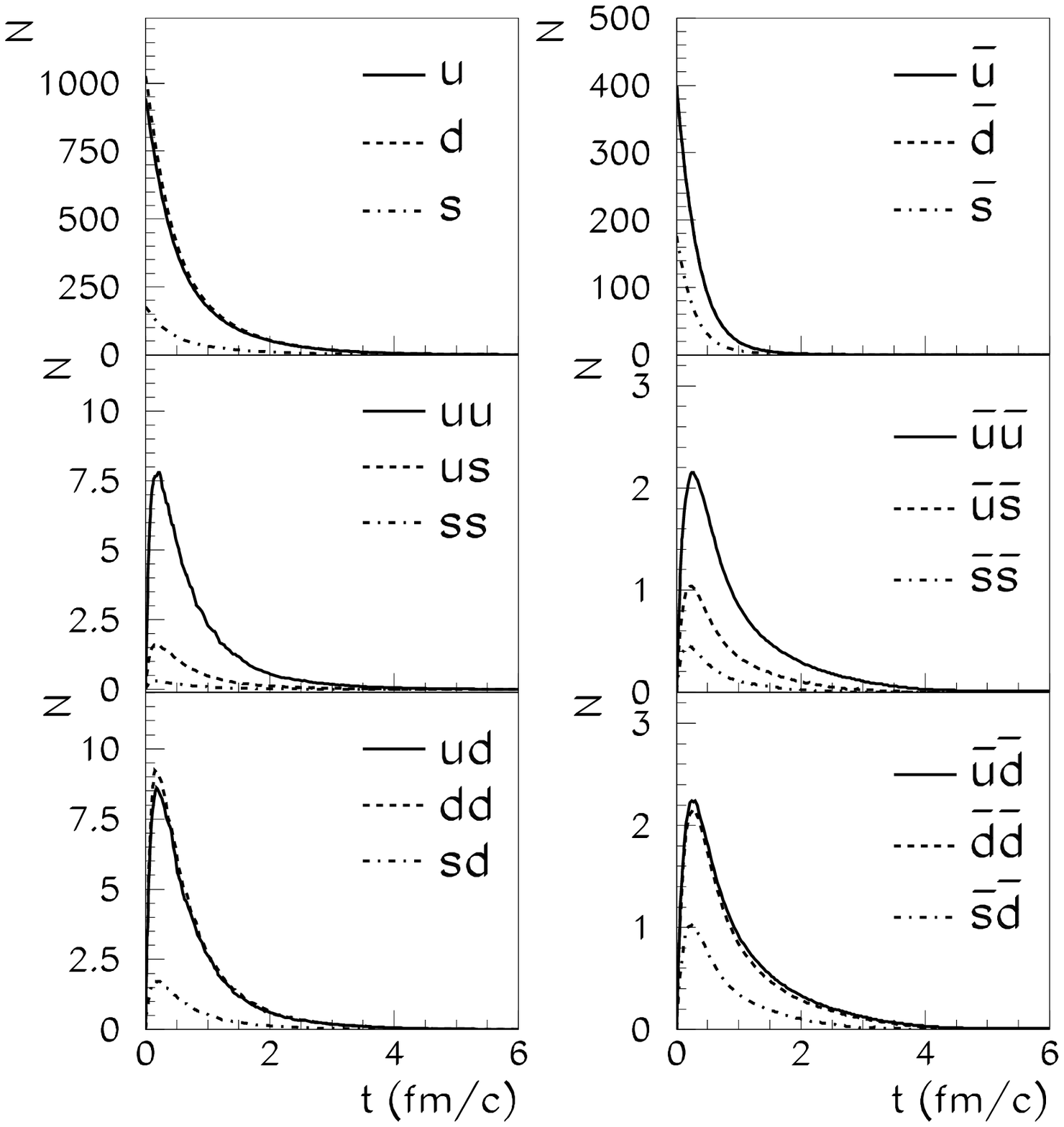}
 \begin{center}
  \begin{minipage}[t]{13.054cm}
  { {\bf Figure 2.}
  The time evolution of the number of colored particles.
  The line styles of different flavor compositions are
  indicated in the respective figures.  }
  \end{minipage}
 \end{center}

\newpage

\insertplot{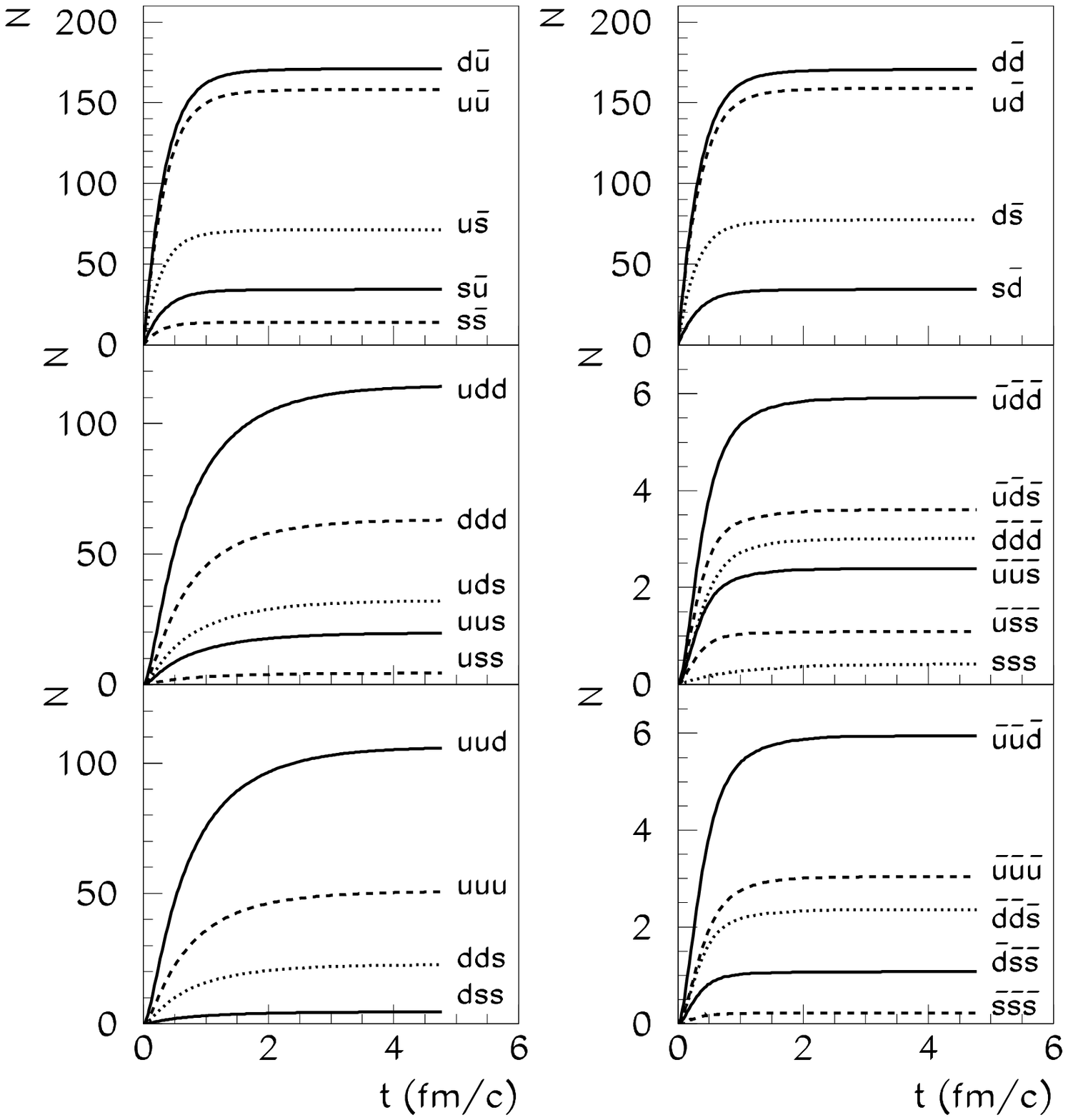}
 \begin{center}
  \begin{minipage}[t]{13.054cm}
  { {\bf Figure 3.}
  The time evolution of the number of color neutral
  clusters consisting of
  quarks and/or anti-quarks indicated on the figures.
 }
  \end{minipage}
 \end{center}

\newpage
\begin{center}
\begin{tabular}{||c||c||c|c|c||}
\hline {\bf Pb+Pb} & { NA49}
& {TrCHEM.}  & { ALCOR } & { RQMD } \\
\hline
\hline
 $h^{-}$  & $680^a$ 
&677.0    &679.8    & \\
\hline
\hline
 $\pi^+$  &  
&581.7    &590.6    &692.9  \\
\hline
 $\pi^0$  &  
&617.0    &605.9    &724.9  \\
\hline
 $\pi^-$  &  
&613.1    &622.0    &728.8  \\
\hline
 $K^+$    & $76^*$  
&\ 79.58  &\ 78.06  &\ 79.0   \\
\hline
 $K^0$    &  
&\ 79.58   &\ 78.06 &\ 79.0  \\
\hline
 ${\overline K}^0$    &  
&\ 39.47  &\ 34.66  &\ 50.4  \\
\hline
 $K^-$    & $\{32\}^b $  
&\ 39.47  &\ 34.66  &\ 50.4   \\
\hline
\hline
 $p^+$    &  
&158.7    &153.2    &199.7  \\
\hline
 $n^0$    &  
&175.8    &170.5    &217.6  \\
\hline
 $\Sigma^+$    &  
&\ \ 8.38 &\ \ 9.16 &\ 12.9  \\
\hline
 $\Sigma^0$    &  
&\ \ 9.79 &\ \ 9.76 &\ 13.1     \\
\hline
 $\Sigma^-$    &  
&\ \ 9.79 &\ 10.39  &\ 13.3    \\
\hline
 $\Lambda^0$   &  
&\ 46.79  &\ 48.85  & \ 35.3  \\
\hline
 $\Xi^0$  &  
&\ \ 4.40 &\ \ 4.89 & \  \ 4.2   \\
\hline
 $\Xi^-$  &  
&\ \ 4.43 &\ \ 4.93 & \ \ 4.2   \\
\hline
 $\Omega^{-}$  &  
&\ \ 0.42 &\ \ 0.62 &   \\
\hline
\hline
 ${\overline p}^-$   &  
&\ \ 8.98 &\ \ 6.24  &\ 27.9    \\
\hline
 ${\overline n}^0$   &  
&\ \ 8.93 &\ \ 6.24  &\ 27.9   \\
\hline
 ${\overline \Sigma}^-$   &  
&\ \ 1.00 &\ \ 0.91   &\ \ 4.6   \\
\hline
 ${\overline \Sigma}^0$   &  
&\ \ 1.09 &\ \ 0.91   &\ \ 4.6    \\
\hline
 ${\overline \Sigma}^+$   &  
&\ \ 0.99 &\ \ 0.91   &\ \ 4.6    \\
\hline
 ${\overline \Lambda}^0$   & 
&\ \ 5.24 &\ \ 4.59  &\ 10.7  \\
\hline
 ${\overline \Xi}^0$   &  
&\ \ 1.08 &\ \ 1.12  &\ \ 2.0    \\
\hline
 ${\overline \Xi}^+$   &   
&\ \ 1.08 &\ \ 1.12  &\ \ 2.0    \\
\hline
 ${\overline \Omega}^{+}$   &  
&\ \ 0.22 &\ \ 0.35  &   \\
\hline
\hline
 $K^0_{S}$   &$ \{54\}^{b,c}$  
&\ 59.66  &\ 56.36  & \ 63.5  \\
\hline
$p^+-{\overline p}^-$ &$\{145\}^a$ 
&149.7    &147.0   &171.8 \\
\hline
 $\Lambda^0$-like &$\{50\pm 10\}^b$  
&\ 56.58  &\ 69.07  & 56.8    \\
\hline
 ${\overline \Lambda}^0$-like &$\{8\pm 1.5\}^b$  
&\ 6.34   &\ 8.12   & 19.3   \\
\hline
\hline
\end{tabular}
\end{center}

\vskip 0.5cm
\noindent {\bf Table 1:}
Total hadron multiplicities for $Pb+Pb$ collision 
at 158 GeV/nucleon bombarding energy. The displayed 
experimental results are from the
NA49 Collaboration: estimated result ${}^a$ is from \cite{NA49QM96};
${}^b$ is from \cite{NA49S97}; ${}^c$ is from 
\cite{NA49S96}; ${}^*$ is estimated from $\{K^-\}$ and
$\{K^0_S\}$. Theoretical results are from  
the Transchemistry, ALCOR
and RQMD  \cite{RQMD} 
("ropes + no rescattering" version) model. Here it is 
$\Lambda^0-{\rm like} \equiv\Lambda^0+\Sigma^0+\Xi^-+\Xi^0+\Omega^-$. 

\newpage

\begin{center}
\begin{tabular}{||c||c|c|c|c||}
\hline {\bf Pb+Pb} &  {\bf WA97}  & {\bf TrCHEM} & {\bf ALCOR} 
		    \\
\hline
\hline
${\overline {\Lambda}}^0/\Lambda^0$
 &$0.128 \pm 0.012$ &0.112&0.117     \\
\hline
${\overline {\Xi}}^+/\Xi^-$
 &$0.266 \pm 0.028$ &0.243&0.227     \\
\hline
${\overline {\Omega}}^+/\Omega^-$
 &$0.46  \pm 0.15$  &0.529&0.564     \\
\hline
$\Xi^-/\Lambda^0$
 &$0.093 \pm 0.007$ &0.078&0.071     \\
\hline
${\overline {\Xi}}^+/{\overline {\Lambda}}^0$
 &$0.195 \pm 0.023$ &0.170&0.138     \\
\hline
$\Omega^-/\Xi^-$
 &$0.195 \pm 0.028$ &0.095&0.125    \\
\hline
\hline
\end{tabular}
\end{center}
\vskip 0.8cm

\noindent {\bf Table 2:}
Strange baryon and anti-baryon ratios measured by
WA97 Collaboration \cite{WA97QM97}
and obtained from Transchemistry and ALCOR model
for $Pb+Pb$ collision at 158 GeV/nucl bombarding energy.
The experimental data are the production ratios in 
mid rapidity at $p_T>0 \ GeV$.

\section*{Acknowledgments}

Stimulating discussions with J.Knoll, A.A.Shanenko, V.Toneev are acknowledged.
This work was supported by the Hungarian Science Fund grants T024094 and
T019700, by the US-Hungarian Science and Technology Joint Fund No. 652/1998,
and by a common project of the Deutsche Forshungsgemeinschaft
and the Hungarian Academy of Science DFG-MTA 101/1998.

\section*{APPENDIX A: \\
THE RELEVANT HYDRODYNAMICAL AND \\ THERMODYNAMICAL EXPRESSIONS}
\label{sec_hydro}

The familiar energy-momentum conservation 
\be
\partial_{\nu}T^{\mu\nu}=0,
\ee
with 
\be
T^{\mu\nu} = (\varepsilon + p) u^{\mu} u^{\nu} - p g^{\mu\nu},
\ee
being the energy-momentum tensor of a perfect fluid with flow
four-velocity $u_{\mu}$, local energy density $\varepsilon$ and
pressure $p$, has two interesting projections: one is parallel
to the four-velocity,
\be
u_{\mu}\partial_{\nu}T^{\mu\nu} =
\partial_{\nu}(\varepsilon u^{\nu}) + p (\partial_{\nu}u^{\nu}) = 0,
\label{ENERG-FLOW}
\ee
and the other is orthogonal to this,
\be
\left( u^{\mu}(u^{\nu}\partial_{\nu}) - \partial^{\mu}\right) p
+ (\varepsilon + p) (u^{\nu}\partial_{\nu}) u^{\mu} = 0.
\label{PRESS-FLOW}
\ee
The former - expressing local energy conservation -
can be casted into the suggestive form
\be
dE+pdV=0
\label{COOL}
\ee
using an infinitesimal volume element, $dV=(\partial_{\mu} u^{\mu})Vd\tau$, 
the total internal energy, $E=\varepsilon V$, 
and the parametric derivation with respect to the time-variable
in the co-moving fluid cell,
\be
\frac{d}{d\tau} = u^{\nu}\partial_{\nu} \ .
\ee

Starting from the first principal theorem of thermodynamics
\be
dE = TdS - pdV + \sum_i \mu_i dN_i,
\label{PRINC}
\ee
and using eq.(\ref{COOL}) we arrive at the following change of
rate of the total entropy:
\be
\dot{S} = - \sum_i \frac{\mu_i}{T} \dot{N}_i,
\ee
where the 'dot' stands for the time derivative ($\dot{S}=dS/dt$).
The chemical potentials $\mu_i$ can generally be derived from
the equation of state.

We consider a flow pattern given by the Bjorken-flow
\be
u_{\mu} = \left( \frac{t}{\tau}, \frac{z}{\tau}, 0, 0 \right). 
\label{FLOW-ANSATZ}
\ee
Here
\ba
\tau^2 &=& t^2 - z^2, \NL
\ea
and the four-velocity is normalized to $u_{\mu}u^{\mu}=1$ and its 
four-divergence is
\be
\partial_{\mu} u^{\mu} = \frac{1}{\tau} \ .
\ee
The energy density $\epsilon$ and the pressure $p$ depend only on
$\tau$ and the flow is stationary. 
Corrections stemming from a mild radial flow also can be worked out, 
but we do not use these corrections here. 
The  effective dimensionality of the flow is $1$ in this case.

Since we deal with a system of rate equations for particle
species numbers $N_i$ occupying a common reaction volume $V$
at temperature $T$, the generating thermodynamical functional
is the free energy, $F(V,T,N_i)$. The chemical potentials are
given by
\be
\mu_i = \frac{\partial F}{\partial N_i},
\ee
the pressure by
\be
p = -  \frac{\partial F}{\partial V},
\ee
the entropy by
\be
S = -  \frac{\partial F}{\partial T},
\ee
and finally the internal energy is
\be
E = F + TS.
\ee

As a starting point we consider a mixture of ideal gases of massive
quarks, diquarks, mesons and baryons, and their respective anti-particles.
The corresponding free energy is
\be
F_{{\rm id}} = \sum_i T N_i \left( \ln \frac{N_i}{N_{i,{\rm th}}} - 1 \right)
\, + \, \sum_i m_iN_i,
\ee
with Maxwell-Boltzmann statistics for the non-relativistic
massive matter:
\be
N_i^{{\rm th}}  =  V d_i \int \! \frac{d^3p}{(2\pi)^3} \,
e^{-p^2/2m_iT}.
\label{MaxBolt}
\ee
Here  $d_i=(2s_i+1)c_i$ are spin and color degeneracy factors.
The chemical potentials in an ideal gas mixture are
\be
\mu_{i,{\rm id}} = T \ln \frac{N_i}{N_{i}^{\rm th}} + m_i.
\ee
The total entropy is given by
\be
S_{{\rm id}} =  \sum_i N_i \left( \frac{5}{2} -  
\ln \frac{N_i}{N_{i}^{\rm th}} \right).
\label{ENTROPY}
\ee
The energy and pressure of such an ideal, non-relativistic mixture
is given by 
\ba
E_{{\rm id}} & = & \sum_i \left( m_i + \frac{3}{2}T \right) N_i, 
\NL
p_{{\rm id}} & = & \sum_i  N_i \frac{T}{V}.
\label{PRESSURE}
\ea


\end{document}